# Strong variation of spin-orbit torques with relative spin relaxation rates in ferrimagnets

Lijun Zhu[1,2*] and Daniel C. Ralph[3,4]

1. State Key Laboratory of Superlattices and Microstructures, Institute of Semiconductors, Chinese Academy of Sciences, Beijing 100083, China
2. College of Materials Science and Opto-Electronic Technology, University of Chinese Academy of Sciences, Beijing 100049, China
3. Cornell University, Ithaca, New York 14850, USA
4. Kavli Institute at Cornell, Ithaca, New York 14850, USA

*ljzhu@semi.ac.cn

**Spin-orbit torques (SOTs) have been widely understood as an *interfacial* transfer of spin that is independent of the bulk properties of the magnetic layer. Here, we report that SOTs acting on ferrimagnetic $Fe_xTb_{1-x}$ layers decrease and vanish upon approaching the magnetic compensation point because the rate of spin transfer to the magnetization becomes slower than the rate of spin relaxation into the crystal lattice due to spin-orbit scattering. These results indicate that the relative rates of competing spin relaxation processes within magnetic layers play a critical role in determining the strength of SOTs, which provides a unified understanding for the diverse and even seemingly puzzling SOT phenomena in ferromagnetic and compensated systems. Our work indicates that spin-orbit scattering within the magnet should be minimized for efficient SOT devices. We also find that the interfacial spin-mixing conductance of interfaces of ferrimagnetic alloys (such as $Fe_xTb_{1-x}$) is as large as that of 3$d$ ferromagnets and insensitive to the degree of magnetic compensation.**

## Introduction

Efficient manipulation of magnetic materials is essential for spintronic devices. While spin-orbit torques (SOTs)[1,2] are well established to be an effective tool to manipulate metallic 3$d$ ferromagnets (FMs), whether they can effectively control antiferromagnetically-ordered systems has remained elusive despite the recent blooming of interest in ferrimagnets (FIMs) and antiferromagnets (AFs)[3-9]. Experimentally, for reasons unclear, the SOTs exerted on nearly compensated FIMs[5-7,10] are often measured to be considerably weaker than those on 3$d$ FMs for a given spin-current generator (by up to >20 times, see below). More strikingly, it remains under debate whether uniform, perfectly compensated FIMs ($M_s$ = 0 emu/cm$^3$) can be switched at all by SOTs [11-13].

Microscopically, SOTs have been widely assumed as an *interfacial* transfer of spin (*i.e.*, spin dephasing length $\lambda_{dp} \approx 0$ nm for transverse spin current) that is independent of the bulk properties of the magnetic layer, such as in drift-diffusion analyses[14-16]. Under this assumption, spin current entering the magnet from an adjacent spin-generating layer is absorbed fully by the magnetization via dephasing to generate SOTs, and the dampinglike SOT efficiency per current density ($\xi_{DL}^j$) will depend *only* on the spin Hall ratio ($\theta_{SH}$) of the spin-generating layer and the spin transparency ($T_{int}$) of the interface which determines what fraction of the spin current enters the magnet [17,18], *i.e.*,

$$\xi_{DL}^j = T_{int}\theta_{SH}. \quad (1)$$

This picture is a reasonable approximation for sufficiently thick metallic FMs that have a short $\lambda_{dp}$ (≤1 nm) due to strong exchange coupling [19-21] and a long spin diffusion length associated with spin relaxation due to spin-orbit scattering[22,23]. However, in antiferromagnetically ordered systems $\lambda_{dp}$ can be quite long, as predicted more than a decade ago[24-27], which, as we discuss below, questions the widely accepted models of "*interfacial torques*", at least, in FIMs and AFMs. So far, any roles of the bulk properties of the magnetic layer, *e.g.*, the competing spin relaxation rates, in the determination of $\xi_{DL}^j$ have been overlooked in SOT analyses.

Here, we report measurements of SOTs acting on ferrimagnetic $Fe_xTb_{1-x}$ layers with strong spin-orbit coupling (SOC)[8] by tuning the $Fe_xTb_{1-x}$ composition and temperature (*T*). We find that, in contrast to the prediction of Eq. (1), $\xi_{DL}^j$ varies strongly with the degree of magnetic compensation for a given $T_{int}$, due to changes in the fraction of spin current that relaxes directly to the lattice via SOC instead of being absorbed by the magnetization to apply SOTs. These results uncover the critical role of spin relaxation rates of the magnetic layer and provide a unified understanding for the diverse SOT phenomena in ferromagnetic and antiferromagnetically ordered systems.

**Sample details**

For this work, we sputter-deposited $Pt_{0.75}Ti_{0.25}$ (5.6 nm)/$Fe_xTb_{1-x}$ (8 nm) bilayers with different Fe volumetric concentrations (*x* = 0.3-1). The $Pt_{0.75}Ti_{0.25}$ layer, a dirty-limit Pt alloy with strong intrinsic spin Hall effect[17], sources spin current that exerts SOT on the FIM $Fe_xTb_{1-x}$ (the spin diffusion length is expected to be ≤ 8 nm at temperatures in this study [22]). Each sample was deposited by co-sputtering on an oxidized Si substrate with a 1 nm Ta seed layer, and protected by a 2 nm MgO and a 1.5 nm Ta layer that was oxidized upon exposure to atmosphere. For electrical measurements, the samples were patterned into 5×60 µm² Hall bars by photolithography and ion milling with a water-cooled stage. After processing, the magnetization hysteresis of the $Fe_xTb_{1-x}$ measured from the anomalous Hall voltage ($V_{AH}$) in patterned Hall bars shows fairly close coercivity (perpendicular depinning field) and squareness as the magnetization of unpatterned regions of the films measured by a superconducting quantum interference device (see Figs. 1(a) and 1(b), more details about the magnetization measurements can be found in Sec. 1 in the Supplementary Materials). As shown in Figs. 1(a)-1(d), the $Fe_xTb_{1-x}$ has strong bulk perpendicular magnetic anisotropy (PMA) for 0.3 ≤ *x* ≤ 0.62 and well-defined in-plane magnetic anisotropy for 0.75 ≤ *x* ≤ 1. All the PMA samples have large anisotropy fields (14.4-72.2 kOe, as estimated from the fits in Fig. S3) and square hysteresis loops for both the out-of-plane magnetization and anomalous Hall voltage.

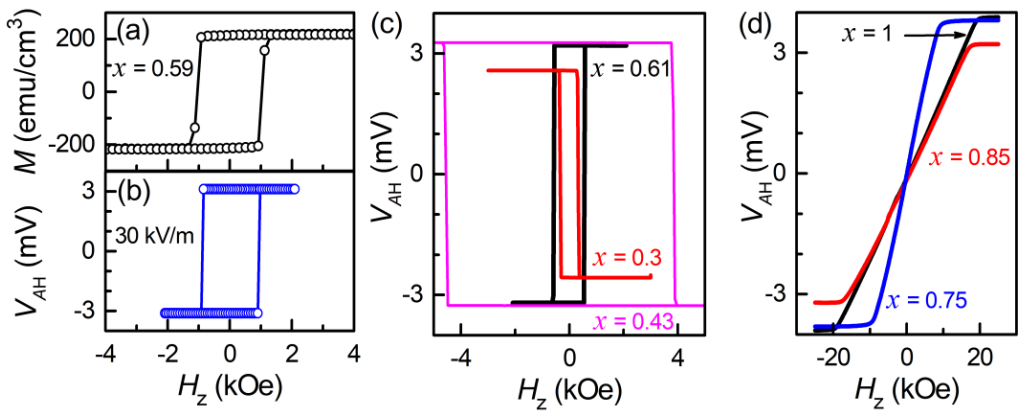

Fig. 1. (a) Magnetization vs out-of-plane field ($H_z$) and (b) Anomalous Hall voltage ($V_{AH}$) vs $H_z$ for $Pt_{0.75}Ti_{0.25}$ (5.6 nm)/$Fe_{0.59}Tb_{0.41}$ (8 nm), indicating strong perpendicular magnetic anisotropy and a high coercivity of ≈1 kOe. $V_{AH}$ vs $H_z$ for $Pt_{0.75}Ti_{0.25}$ (5.6 nm)/$Fe_xTb_{1-x}$ (8 nm) with (c) perpendicular (*x* = 0.3, 0.43, and 0.61) and (d) in-plane magnetic anisotropy (*x* = 0.75, 0.85, and 1).

**Strong Variation of Spin-Orbit Torques with Composition and temperature**

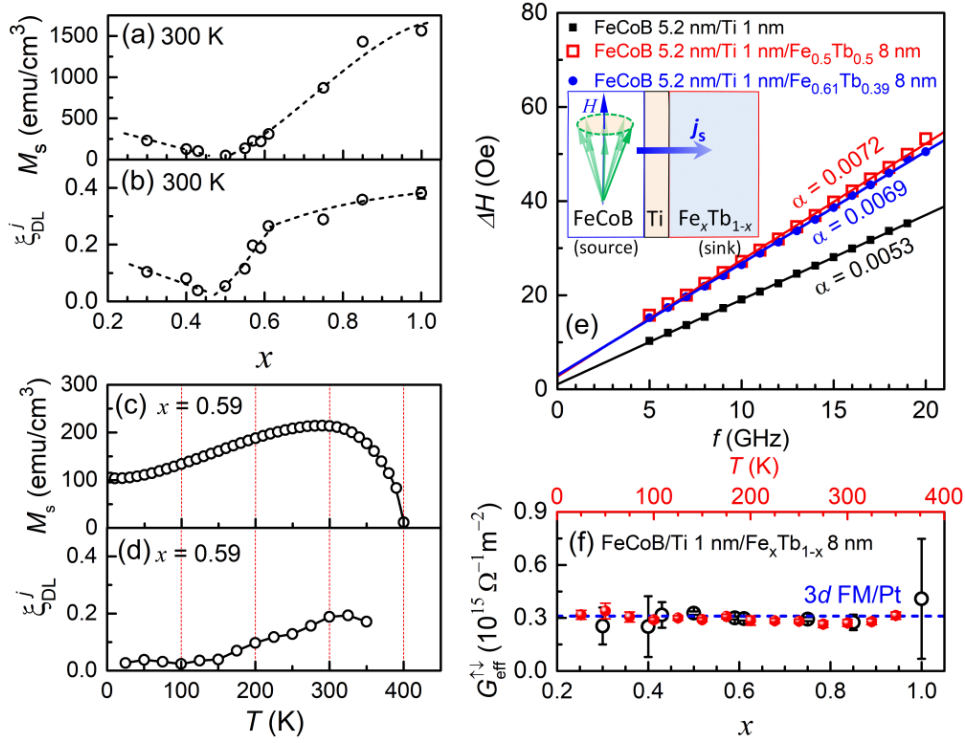

Fig. 2. (a) $M_s$ and (b) $\xi^j_{DL}$ for $Pt_{0.75}Ti_{0.25}/Fe_xTb_{1-x}$ with different Fe concentration ($x$) at 300 K. (c) $M_s$ and (d) $\xi^j_{DL}$ for $Pt_{0.75}Ti_{0.25}/Fe_{0.59}Tb_{0.41}$ at different temperatures. (e) Frequency dependence of ferromagnetic resonance linewidth ($\Delta H$) of the FeCoB layer in FeCoB (5.2 nm)/Ti (1 nm), FeCoB (5.2 nm)/Ti (1 nm)/$Fe_{0.5}Tb_{0.5}$ (8 nm), and FeCoB (5.2 nm)/Ti (1 nm)/$Fe_{0.61}Tb_{0.39}$ (8 nm) samples. The solid lines represent linear fits, the slopes of which yield the damping. In (a)-(e) some error bars are smaller than the data points. (f) $G^{\uparrow\downarrow}_{eff}$ of the FeCoB/Ti/$Fe_xTb_{1-x}$ interfaces measured from spin pumping into the $Fe_xTb_{1-x}$. The blue circles are for the composition series (300 K) and the red dots for the temperature series ($x = 0.59$). The blue dashed line represents $G^{\uparrow\downarrow}_{eff} = 0.31 \times 10^{15}$ $\Omega^{-1}$ $m^{-2}$ previously reported for typical Pt/3d FM interfaces [41].

We performed harmonic Hall voltage response (HHVR) measurements[37,38] by carefully separating out any thermoelectric effects (see details in Sec. 1 in the Supplementary Materials). We calculate the SOT efficiency using $\xi^j_{DL} = (2e/\hbar) M_s t_{FeTb} H_{DL}/j_c$, [18] where $e$ is elementary charge, $\hbar$ reduced Plank's constant, $t_{FeTb}$ the $Fe_xTb_{1-x}$ thickness, $M_s$ the saturation magnetization of the $Fe_xTb_{1-x}$ (see Sec. 2 in in the Supplementary Materials), and $j_c$ the current density in the $Pt_{0.75}Ti_{0.25}$. $H_{DL}$ is the current-driven damping-like SOT field. The "planar Hall correction" [32,33,38-40] is negligible for the PMA $Fe_xTb_{1-x}$ samples ($V_{Ph}/V_{AH}<0.1$, see Fig. S4 in in the Supplementary Materials).

In Figs. 2(a) and 2(b) we show the measured values of $M_s$ and $\xi^j_{DL}$ at 300 K for the $Pt_{0.75}Ti_{0.25}/Fe_xTb_{1-x}$ bilayers with different $Fe_xTb_{1-x}$ compositions (we refer to this as the composition series). $M_s$ decreases monotonically by a factor of 33, from 1560 emu/cm³ for $x = 1$ (pure Fe, 3d FM) to 47 emu/cm³ for $x = 0.5$ (nearly full compensation), and then increases slowly as $x$ further decreases. More details about the composition dependent magnetic properties are shown in Sec. 2 in in the Supplementary Materials. As $x$ decreases in the Fe-dominated regime ($x \geq 0.5$), $\xi^j_{DL}$ decreases by a factor of 7 at 300 K, first slowly from $0.38 \pm 0.02$ for $x = 1$ to $0.27 \pm 0.01$ for $x = 0.61$ and then more rapidly to 0.054 ± 0.002 for $x = 0.5$. $\xi_{DL}^j$ increases slowly with decreasing $x$ in the Tb-dominated regime ($x < 0.5$). The fieldlike SOT from the same HHVR measurements is smaller than $\xi_{DL}^j$ for each $x$ and also varies with $x$ (Fig. S8). We also measured $M_s$ and $\xi_{DL}^j$ of $Pt_{0.75}Ti_{0.25}/Fe_{0.59}Tb_{0.41}$ as a function of temperature (we refer to this as the temperature series). Upon cooling from 350 K to 25 K, $M_s$ and $\xi_{DL}^j$ for the $Pt_{0.75}Ti_{0.25}/Fe_{0.59}Tb_{0.41}$ sample are tuned by > 2 times (Fig. 2(c)) and by > 7.5 times (Fig. 2(d)), respectively. The dramatic tuning of $\xi_{DL}^j$ by the $Fe_xTb_{1-x}$ composition and temperature is a striking observation because it suggests a strong dependence of SOTs on some bulk properties of the $Fe_xTb_{1-x}$, in contrast to $\xi_{DL}^j$ for heavy metal (HM)/3d FM samples which is insensitive to the type of the FM [41] and temperature[42].

**Robustness of the Spin Hall ratio and the effective spin-mixing conductance**

These strong variations cannot be explained by changes in either $\theta_{SH}$ or $T_{int}$ (Eq. (1)). First, $\theta_{SH}$ is a property of the $Pt_{0.75}Ti_{0.25}$ layer, not the $Fe_xTb_{1-x}$ layer. The $Pt_{0.75}Ti_{0.25}$ layer is made identically for all of the samples, and has a sufficiently large resistivity ($\rho_{xx}$ = 135 μΩ cm) such that its properties can hardly be affected significantly by either a neighboring layer or temperature. We have verified that $\xi_{DL}^j$ for a ferromagnetic $Pt_{0.75}Ti_{0.25}$ (5.6 nm)/FePt (8 nm) bilayer only has very weak temperature dependence (Fig. S14 in the Supplementary Materials), in good consistence with previous reports on other HM/3d FM samples[42]. We have also measured negligible SOT signal from the 8 nm $Fe_xTb_{1-x}$ layers in control samples without a $Pt_{0.75}Ti_{0.25}$ layer (Sec. 7 in the Supplementary Materials), indicating that changes in our signals are not due to SOT arising from the $Fe_xTb_{1-x}$ bulk. Note that a bulk torque of a magnetic layer is strongly thickness dependent[43,44] and vanishes at small thicknesses of a few nm [43].

As for the possibility of changes in $T_{int}$, if we employ a drift-diffusion analysis [14-16], the effect on $T_{int}$ of spin backflow at the $Pt_{0.75}Ti_{0.25}/Fe_xTb_{1-x}$ interface should be proportional to the effective spin-mixing conductance ($G_{eff}^{\uparrow\downarrow}$) of the interface, i.e. $T_{int} \approx 2G_{eff}^{\uparrow\downarrow}/G_{PtTi}$,[45] with $G_{PtTi} = 1/\lambda_s\rho_{xx} \approx 1.3\times10^5$ $\Omega^{-1}$ m$^{-1}$ [18,46] being the spin conductance of the $Pt_{0.75}Ti_{0.25}$. To quantify $G_{eff}^{\uparrow\downarrow}$, we measure the change in the damping ($\alpha$) of a precessing $Fe_{60}Co_{20}B_{20}$ (= FeCoB) layer due to the absorption of the FeCoB-emitted spin current at the $Fe_xTb_{1-x}$ interfaces (Fig. 2(e) and Fig. S9). The samples used here had the structure FeCoB (5.2 nm)/Ti (1 nm) and FeCoB (5.2 nm)/Ti (1 nm)/$Fe_xTb_{1-x}$ (8 nm). Each of these samples is sputter-deposited on a 1 nm Ta seed layer and protected by capping with MgO (2 nm)/Ta (1 nm). The value of $\alpha$ for the FeCoB layers is determined from the linear dependence of the ferromagnetic resonance linewidth ($\Delta H$, half width at half maximum) on the frequency ($f$) using the relation $\Delta H = \Delta H_0 + 2\pi\alpha f/\gamma$, where $\Delta H_0$ is the inhomogeneous broadening of the linewidth and $\gamma$ the gyromagnetic ratio. The damping enhancement of the FeCoB layer induced by spin pumping into the 8 nm $Fe_xTb_{1-x}$ layers, $\Delta\alpha = \alpha_{FeCoB/Ti/FeTb} - \alpha_{FeCoB/Ti}$, can be related to $G_{eff}^{\uparrow\downarrow}$ as [47-49]

$$\Delta\alpha = \gamma\hbar^2 G_{eff}^{\uparrow\downarrow}/2e^2 M_{FeCoB} t_{FeCoB} \qquad (2)$$

where $t_{FeCoB}$ = 5.2 nm and $M_{FeCoB}$ = 1255 emu/cm$^3$ is the saturation magnetization of the FeCoB layer as measured by SQUID. The value of $\alpha$ = 0.0053 for the bare FeCoB/Ti sample with no $Fe_xTb_{1-x}$ coincides closely with the intrinsic damping of FeCoB ($\approx 0.006$[41]), indicating that the damping in this system does not contain any significant contributions from interfacial two-magnon scattering or spin memory loss. As shown in Fig. 2(f), $G_{eff}^{\uparrow\downarrow}$ of the FeCoB/Ti/$Fe_xTb_{1-x}$ interfaces is insensitive to temperature and the $Fe_xTb_{1-x}$ composition within experimental uncertainty, and has a value as high as that of typical 3d FM/Pt interfaces ($\approx 0.31\times10^{15}$ $\Omega^{-1}$ m$^{-2}$) [41]. This indicates that compensated Fe$_x$Tb$_{1-x}$ alloys act as spin sinks that are just as good as 3$d$ FMs and Pt, and that there is no enhancement in the amount of spin reflection and backflow due to magnetic compensation. In principle, changes in $T_{int}$ for SOT measurements could also arise from spin memory loss induced by interfacial SOC[38,50,51], but this should be a minor effect for $T_{int}$ of our un-annealed Pt$_{0.75}$Ti$_{0.25}$/Fe$_x$Tb$_{1-x}$ just as is the case of un-annealed Pt/Co [52]. As noted above, we also do not observe any enhancement in damping due to spin memory loss in the spin-pumping measurements.

**Variation of SOT with the relative spin relaxation rates**

Since we have ruled out any significant change in $\theta_{SH}$ or $T_{int}$ as contributing to the large variations we measure in $\xi_{DL}^j$ as a function of composition and temperature, these large variations must be due to physics that is not captured in the simple Eq. (1). We suggest that spin relaxation induced by SOC in the bulk of the Fe$_x$Tb$_{1-x}$ layer is the most likely physics that is neglected in Eq. (1). As schematically shown in Fig. 3(a), a spin current in general can be relaxed in a magnetic layer through two competing mechanisms: exchange-interaction-induced angular momentum transfer from spin current to magnetization (with a relaxation rate $\tau_M^{-1}$) and bulk spin-orbit-scattering-induced loss of spin angular momentum to the lattice (with a relaxation rate $\tau_{so}^{-1}$). Theory [24-27] and experiments [53,54] have suggested that, in fully or partially compensated magnetic systems, the rate of spin angular momentum transfer via exchange interaction can be greatly decreased because of cancellations between exchange fields of antiferromagnetically aligned magnetic sub-lattices, resulting in long $\lambda_{dp}$ and low $\tau_M^{-1}$. This makes it possible for spin currents in ferrimagnets to relax partially or even primarily via spin-orbit scattering to the lattice, instead of applying a spin-transfer torque to the magnetization.

We can consider how the SOT should scale as a function of the ratio $\tau_M^{-1}/\tau_{so}^{-1}$. Quantitative measurements of these rates (e.g., from the dependence on layer thicknesses of spin valve or spin-pumping experiments) are quite challenging because the bulk properties of Fe$_x$Tb$_{1-x}$ [55] and other ferrimagnetic alloys[53,54] vary sensitively with the layer thicknesses [53-55](*e.g*., the magnetic compensation, the bulk PMA, the orientation of the magnetic easy axis, and resistivity all change with thickness). Nonetheless, it is reasonable to expect $\tau_M^{-1} \propto M_s$ for such ferrimagnetic alloys considering the cancelling effects of the exchange fields from the antiferromagnetically-aligned magnetic sub-lattices. For the spin-orbit scattering rate, the Elliot-Yafet mechanism [56-58] predicts $\tau_{so}^{-1} \propto \zeta_{so}\tau_e^{-1}$, where $\zeta_{so}$ is the bulk SOC strength and $\tau_e^{-1}$ is the momentum scattering rate. One can thus expect $\tau_M^{-1}/\tau_{so}^{-1} = kM_s/\zeta_{so}\tau_e^{-1}$, with $k$ being a constant. Since only the spin current relaxed through exchange interaction with magnetization contributes to SOTs, we propose that

$$\xi_{DL}^j \approx \xi_{DL,0}^j \ \tau_M^{-1}/(\tau_M^{-1} + \tau_{so}^{-1}) = \xi_{DL,0}^j(1+(kM_s/\zeta_{so}\tau_e^{-1})^{-1})^{-1}, (3)$$

where $\xi_{DL,0}^j$ is the value of $\xi_{DL}^j$ in the limit $\tau_M^{-1}/\tau_{so}^{-1} \gg 1$.

Figures 3(b)-3(d) show the estimated values of $\tau_e^{-1}$, $\zeta_{so}$, and $\zeta_{so}\tau_e^{-1}$ for both our composition series ($x$ = 0.3-1, $T$ = 300 K, black symbols) and our temperature series ($x$ = 0.59, $T$ = 25-350 K, red symbols). Here, $\tau_e^{-1}$ is estimated from the resistivity of the Fe$_x$Tb$_{1-x}$ following $\rho_{FeTb} = m^*/ne^2\tau_e$, where $m^* \approx 9.1 \times 10^{-31}$ kg is an effective mass of carriers and $n$ is the carrier density measured from the ordinary Hall coefficient ($R_{OH} = 1/ne$ for a single-band model [59], see Sec. 6 in the Supplementary Materials). $\tau_e^{-1}$ of the Fe$_x$Tb$_{1-x}$ varies by a factor of 2 by composition and a factor of 6 by temperature, suggesting a significant tuning of the Fermi surface properties. The average SOC strength of the Fe$_x$Tb$_{1-x}$ is estimated as $\zeta_{so} \approx x\zeta_{so,Fe} + (1-x)\zeta_{so,Tb}$ using the theoretical values of $\zeta_{so,Fe}$ = 0.069 eV for Fe and $\zeta_{so,Tb} = 0.283$ eV for Tb[60]. We estimate that $\zeta_{so}\tau_e^{-1}$ decreases by a factor of > 16 as $x$ varies between 0.3 to 1 and by a factor of > 7 as temperature increases from 25 K to 350 K (Fig. 3(d)). In Fig. 3(e) we plot $\xi_{DL}^j$ as a function of $M_s/\zeta_{so}\tau_e^{-1}$ for both the composition series and the temperature series. As $M_s/\zeta_{so}\tau_e^{-1}$, or equivalently $\tau_M^{-1}/\tau_{so}^{-1}$, decreases, we find that $\xi_{DL}^j$ decreases first slowly and then rapidly towards a vanishing value[61]. The variation of $\xi_{DL}^j$ with $M_s/\zeta_{so}\tau_e^{-1}$ can be fit very well by Eq. (3) with $\xi_{DL,0}^j = 0.395 \pm 0.022$ and $k = (1.66 \pm 0.23)\times 10^{11}$ s$^{-1}$ emu$^{-1}$ cm$^3$ eV.

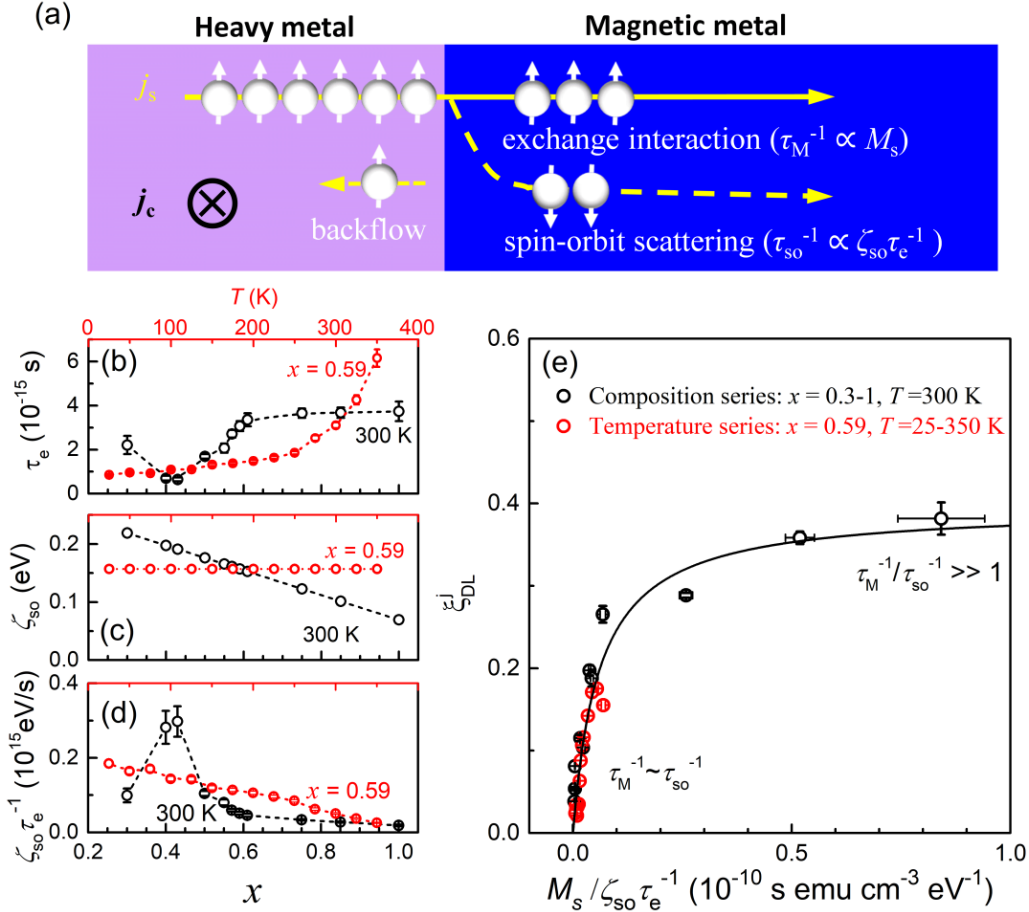

Fig. 3. (a) Schematic of the spin relaxation processes that can influence the SOT, highlighting the competition between exchange interaction (with relaxation rate $\tau_M^{-1} \propto M_s$) and spin-orbit scattering ($\tau_{so}^{-1} \propto \zeta_{so}\tau_e^{-1}$). Only the spin current relaxed by exchange interaction contributes to SOTs. (b) Momentum scattering time ($\tau_e$), (c) Estimated SOC strength ($\zeta_{so}$), (d) $\zeta_{so}\tau_e^{-1}$, and (e) $\xi_{DL}^j$ of Pt$_{0.75}$Ti$_{0.25}$/Fe$_x$Tb$_{1-x}$ vs $M_s/\zeta_{so}\tau_e^{-1}$ for the composition series ($x = 0.3$-1, $T = 300$ K, black circles) and for the temperature series ($x = 0.59$, $T = 25$-300 K, red circles). The solid curve in (e) represents the fit of the data to Eq. (3).

Since the strong variation of $\xi_{DL}^j$ with relative spin relaxation rates we propose here is unlikely to be specific just to the HM/Fe$_x$Tb$_{1-x}$ system (as indicated by the general fact that the SOT provided by a given spin-current generator is significantly weaker on FIMs than on FMs, see Table 1), we generalize Eq. (1) as

$$\xi_{DL}^j = T_{int}\theta_{SH}\,\tau_M^{-1}/(\tau_M^{-1} + \tau_{so}^{-1}). \quad (4)$$

Only in FMs and FIMs, which have relatively low resistivity, small SOC, and high-magnetization so that $\tau_M^{-1}/\tau_{so}^{-1} \gg 1$, should the simple form of Eq. (1) apply. In *perfectly* compensated FIMs ($M_s = 0$ emu/cm$^3$) in which $\tau_M^{-1}$ goes to zero, we expect $\xi_{DL}^j$ to go to zero as well. In FIMs or "AF" domains that are only *partially* uncompensated ($M_s > 0$ emu/cm$^3$)[62-65], $\tau_M^{-1}$ can be comparable to $\tau_{so}^{-1}$ so that $\xi_{DL}^j$ is reduced, but not zero.

We note that our conclusions are contrary to some previous experiments which reported $\xi_{DL}^j$ to remain constant[66-68] or even diverge[69] near the magnetic compensation point of HM/CoTb or HM/CoFeGd bilayers. While it might be possible that $\tau_M^{-1}/\tau_{so}^{-1}$ is different in CoTb and CoFeGd compared to Fe$_x$Tb$_{1-x}$ (e.g. Gd has zero atomic orbital angular momentum[8] and thus considerably weaker SOC than Tb), we also question these previous conclusions for a variety of technical experimental reasons. In three of the previous experiments, the PMA of the FIM layer was weak and showed gradual magnetization hysteresis[68], non-parabolic first-harmonic signal in HHVR measurements[66,68,69], and/or non-linear second-harmonic signal in HHVR measurements[66,68] as a function of a small in-plane applied magnetic field. This indicates magnetization behavior outside of the simple macrospin model assumed in the HHVR analysis. References 68,69 also applied "planar Hall correction" to their HHVR results; this often causes erroneous estimates of $\xi_{DL}^j$ (see Tab. 1 for a few examples and refs. 32,33,38,39,40 for more discussions). Reference [66,67] reported substantial changes of sample properties before and after device patterning, resulting in large uncertainties in the estimation of $M_s$ and $\xi_{DL}^j$ for their Hall-bar samples. Reference 68 studied CoTb layers with thicknesses (1.7-2.6 nm) much thinner than the likely spin dephasing length ($\approx$ 10 nm[53]) so that the escape rate from the film was likely faster than either $\tau_M^{-1}$ or $\tau_{so}^{-1}$.

**Table 1**. The out-of-plane HHVR results of the PMA HM/FM samples without and with the "planar Hall correction" vs the in-plane HHVR results on in-plane magnetized samples with the similar HM resistivities and the same FM layer. Applying a large "planar Hall correction" gives unrealistic numbers for the fieldlike and/or dampinglike torque efficiencies and alters the sign of the dampinglike torque of the Pd 4/Co 0.64 and the sign of both dampinglike and fieldlike torque of the W 2.5/CoFeB 1. The PMA results for $\xi_{DL}^j$ are in good agreement with in-plane HHVR results only if the "correction" is not applied.

| PMA samples | $V_{PH}/V_{AH}$ | PMA sample No "correction" | | PMA sample with "correction" | | In-plane sample | | Reference |
|---|---|---|---|---|---|---|---|---|
| | | $\xi_{DL}^j$ | $\xi_{FL}^j$ | $\xi_{DL}^j$ | $\xi_{FL}^j$ | $\xi_{DL}^j$ | $\xi_{FL}^j$ | |
| W 2.2/CoFeB 1 | 0.486 | -0.132 | -0.064 | -3.52 | -3.25 | - | - | [33] |
| W 2.5/CoFeB 1 | 0.54 | -0.15 | -0.005 | 0.93 | 1.00 | - | - | [33] |
| Pt 4/Co 0.75 | 0.31 | 0.21 | -0.049 | 0.29 | 0.13 | 0.19 | -0.046 | [38] |
| Pd 4/Co 0.64 | 0.56 | 0.07 | -0.050 | -0.1 | -0.16 | 0.06 | -0.0002 | [32] |
| Au$_{0.25}$Pt$_{0.75}$ 4/Co 0.8 | 0.33 | 0.30 | -0.12 | 0.39 | -0.14 | 0.32 | -0.020 | [38] |

**Scientific implications**

Taking into account the relative rates of spin-orbit-induced relaxation to the lattice versus spin transfer to the magnetization can resolve outstanding puzzles in previous experiments, and has other important implications. First, the condition $\tau_M^{-1} \ll \tau_{so}^{-1}$ can explain why the measured $\xi_{DL}^j$ values for SOT acting on nearly-compensated FIMs are often several to over 20 times smaller than for corresponding measurements using 3$d$ FMs (see Tab. 2 for a few representative examples with spin current sources that have similar resistivities, thicknesses, and thus similar values of $\theta_{SH}$ and $T_{int}$). We have also verified from the variation of measured $\xi_{DL}^j$ values of a large number of samples that $\tau_M^{-1}/(\tau_M^{-1}+\tau_{so}^{-1}) = 0.58$ for $Co_{0.65}Tb_{0.35}$ layers such that Pt-X/$Co_{0.65}Tb_{0.35}$ is only 58% of that of Pt-X/Co for given Pt-X (Pt-X being Pt-based alloys and multilayers, the detail of which will be published elsewhere). Moreover, the diminishment of SOTs in fully-compensated systems explains the absence of current-induced switching of some HM/AF [11-13], while the sizable SOTs in nearly but not fully compensated systems explains the occurrence of switching of uncompensated "antiferromagnetic" domains by in-plane current [62-65]. Our results suggest that in general SOTs will be reduced in any experiments which use magnetic free layers for which $\tau_{so}^{-1}$ is not much less than $\tau_M^{-1}$. Spin-orbit scattering within the magnetic layer should therefore be minimized and the average exchange coupling maximized for efficient SOT devices. Finally, it will be essential to modify spin transport models to include spin decoherence by spin-orbit scatting on an equal footing with dephasing by the exchange interaction.

**Table 2**. Comparison of $\xi_{DL}^j$ for FIMs and 3d FMs in contact with spin current sources that have similar resistivities, thicknesses, and thus similar values of $\theta_{SH}$ and $T_{int}$.

| spin current source | $\xi_{DL}^j$ | | ratio |
|---|---|---|---|
| | FIM | 3d FM | |
| Ta | -0.03 (CoTb)[5] | -0.12 (FeCoB)[10] | 4 |
| W | -0.04 (CoTb)[70] | -0.44 (FeCoB)[71] | 11 |
| Pt | 0.017 (CoTb)[5] | 0.15 (Co,FeCoB)[10] | 8.8 |
| Pt/NiO | 0.09 (CoTb)[6] | 0.6 (FeCoB)[72] | 6.7 |
| $Pt_{0.75}Ti_{0.25}$ | 0.05 ($Fe_{0.5}Tb_{0.5}$) | 0.38 (Fe) | 7.6 |
| $Bi_2Se_3$ | 0.13 (GdFeCo)[7] | 3.5 (NiFe)[73] | 27 |

**Conclusions**

In summary, we have shown that the strength of SOTs depends critically on the ratio of rate of spin-orbit-induced spin relaxation within a magnetic layer relative to the rate of exchange-induced spin transfer to the magnetization. We find experimentally that SOT efficiencies decrease strongly upon approaching the magnetic compensation point in ferrimagnetic $Fe_xTb_{1-x}$ due to a decrease in the rate of exchange-induced spin transfer on account of partial cancellation between the oppositely-directed exchange interactions from the magnetic sub-lattices. Near the compensation point, spin-orbit-induced spin relaxation dominates over spin transfer to the magnetization so that the measured SOT goes to zero. These results suggest the breakdown of the "*interfacial torques*" concept in FIMs and AFs. We find no indication of any dependence of the spin transparency of $Fe_xTb_{1-x}$ interfaces on the degree of compensation. This work provides not only a unified understanding of the very different efficiencies of SOTs that have been reported in the literature for FMs, FIMs, and AFs, but also insight about how the different sources of spin relaxation should be optimized in the design of FIMs and AFs for spintronic technologies [8,9].

## Acknowledgement

We thank Dahai Wei, Xin Lin, Qianbiao Liu for help with sample deposition. We also thank Tianxiang Nan, Yanan Chai, Wei Han, Liangliang Guo for help with ferromagnetic resonance measurements. This work was supported in part by the Strategic Priority Research Program of the Chinese Academy of Sciences (XDB44000000), in part by the Office of Naval Research (N00014-19-1-2143), in part by the NSF MRSEC program (DMR-1719875) through the Cornell Center for Materials Research, and in part by the NSF (NNCI-2025233) through the Cornell Nanofabrication Facility/National Nanotechnology Coordinated Infrastructure.
## Data availability

The data that support this study are available from the corresponding authors upon reasonable request.

## Conflict of Interest

The authors declare no conflict of interest.

## References

[1] I. M. Miron, K. Garello, G. Gaudin, P.-J. Zermatten, M. V. Costache, S. Auffret, S. Bandiera, B. Rodmacq, A. Schuhl, P. Gambardella, Perpendicular switching of a single ferromagnetic layer induced by in-plane current injection, Nature **476**, 189 (2011).

[2] L. Liu, C. F. Pai, Y. Li, H. W. Tseng, D. C. Ralph, R. A. Buhrman, Spin-Torque Switching with the Giant Spin Hall Effect of Tantalum, Science **336**, 555 (2012).

[3] L. Caretta, M. Mann, F. Büttner, K. Ueda, B. Pfau, C. M. Günther, P. Hessing, A. Churikova, C. Klose, M. Schneider, D. Engel, C. Marcus, D. Bono, K. Bagschik, S. Eisebitt, and G. S. D. Beach, Fast current-driven domain walls and small skyrmions in a compensated ferrimagnet. Nat. Nanotechnol. **13**, 1154 (2018).

[4] K. Cai, Z. Zhu, J. M. Lee, R. Mishra, L. Ren, S. D. Pollard, P. He, G. Liang, K. L. Teo, and H. Yang, Ultrafast and energy-efficient spin–orbit torque switching in compensated ferrimagnets. Nat. Electron. **3**, 37 (2020).

[5] J. Han, A. Richardella, S. A. Siddiqui, J. Finley, N. Samarth, and L. Liu, Room-Temperature Spin-Orbit Torque Switching Induced by a Topological Insulator, Phys. Rev. Lett. **119**, 077702 (2017).

[6] H. Wang, J. Finley, P. Zhang, J. Han, J. T. Hou, and L. Liu, Spin-Orbit-Torque Switching Mediated by an Antiferromagnetic Insulator, Phys. Rev. Applied **11**, 044070 (2019).

[7] H. Wu, Y. Xu, P. Deng, Q. Pan, S. A. Razavi, K. Wong, L. Huang, B. Dai, Q. Shao, G. Yu, X. Han, J.-C. Rojas-Sánchez, S. Mangin, K. L. Wang, Spin-Orbit Torque Switching of a Nearly Compensated Ferrimagnet by Topological Surface States, Adv. Mater. **31**, 1901681 (2019).

[8] F. Radu, and J. Sánchez-Barriga, Ferrimagnetic Heterostructures for Applications in Magnetic Recording. Novel Magnetic Nanostructures, Elsevier, Amsterdam, The Netherlands, 267–331(2018).

[9] Z. Zhang, Z. Zheng, Y. Zhang, J. Sun, K. Lin, K. Zhang, X. Feng, L. Chen, and J. Wang, 3D Ferrimagnetic Device for Multi-Bit Storage and Efficient In-Memory Computing, IEEE Electron Device Letters, **42**, 152-155 (2021).

[10] C.-F. Pai, M. Mann, A. J. Tan, and G. S. D. Beach, Determination of spin torque efficiencies in heterostructures with perpendicular magnetic anisotropy, Phys. Rev. B **93**, 144409 (2016).

[11] Q. Ma, Y. Li, Y.-s. Choi, W.-C. Chen, S. J. Han, and C. L. Chien, Spin orbit torque switching of synthetic Co/Ir/Co trilayers with perpendicular anisotropy and tunable interlayer coupling, Appl. Phys. Lett. **117**, 172403 (2020).

[12] C. C. Chiang, S. Y. Huang, D. Qu, P. H. Wu, and C. L. Chien, Absence of Evidence of Electrical Switching of the Antiferromagnetic Néel Vector, Phys. Rev. Lett. **123**, 227203 (2019).

[13] P. Zhang, J. Finley, T. Safi, and L. Liu, Quantitative Study on Current-Induced Effect in an Antiferromagnet Insulator/Pt Bilayer Film, Phys. Rev. Lett. **123**, 247206 (2019).

[14] P. M. Haney, H. W. Lee, K. J. Lee, A. Manchon, M. D. Stiles, Current induced torques and interfacial spin-orbit coupling: Semiclassical modeling, Phys. Rev. B **87**, 174411 (2013).

[15] Y.-T. Chen, S. Takahashi, H. Nakayama, M. Althammer, S. T. B. Goennenwein, E. Saitoh, and G. E. W. Bauer, Theory of spin Hall magnetoresistance, Phys. Rev. B 87, 144411 (2013).

[16] V. P. Amin and M. D. Stiles, Spin transport at interfaces with spin-orbit coupling: Phenomenology, Phys. Rev. B **94**, 104420 (2016).

[17] L. Zhu, D. C. Ralph, R. A. Buhrman, Maximizing Spin-orbit Torque generated by the spin Hall effect of Pt, Appl. Phys. Rev. **8**, 031308 (2021).

[18] M.-H. Nguyen, D. C. Ralph, and R. A. Buhrman, Spin Torque Study of the Spin Hall Conductivity and Spin
9